\documentclass[superscriptaddress]{revtex4}
\usepackage{lettrine}
\usepackage{amsmath}
\usepackage{mathrsfs}
\usepackage{graphicx}
\usepackage{bm}
\usepackage{amsfonts,dsfont}

\begin{document}
\title{Long-range Cooper pair splitter with high entanglement production rate}
\author{Wei Chen}
\email{weichenphy@nuaa.edu.cn}
\affiliation{College of Science, Nanjing University of Aeronautics and Astronautics, Nanjing 210016, China}
\author{D. N. Shi}
\affiliation{College of Science, Nanjing University of Aeronautics and Astronautics, Nanjing 210016, China}
\author{D. Y. Xing}
\affiliation{National Laboratory of Solid State Microstructures and Department of Physics, Nanjing University, Nanjing 210093, China}
\begin{abstract}
\textbf{Cooper pairs in the superconductor are a natural source of spin entanglement. The existing proposals of the Cooper pair splitter can only realize a low efficiency of entanglement production, and its size is constrained by the superconducting coherence length. Here we show that a long-range Cooper pair splitter can be implemented in a normal metal-superconductor-normal metal (NSN) junction by driving a supercurrent in the S. The supercurrent results in a band gap modification of the S, which significantly enhances the crossed Andreev reflection (CAR) of the NSN junction and simultaneously quenches its elastic cotunneling. Therefore, a high entanglement production rate close to its saturation value can be achieved by the inverse CAR. Interestingly, in addition to the conventional entangled electron states between opposite energy levels, novel entangled states with equal energy can also be induced in our proposal.}
\end{abstract}
\maketitle

\lettrine{G}{}eneration and detection of the electron entanglement in solid state physics have attracted great scientific interest, for the prospect of large-scale implementation of quantum information and computation schemes \cite{Nielsen}. The conventional BCS superconductor are considered as a natural source of spin entanglement \cite{Recher,Lesovik}, for the Cooper pair in the superconductor can break up into two nonlocal entangled electrons that enter into different normal terminals via the crossed Andreev reflection (CAR) \cite{Byers,Feinberg}. Recently, the feasibility of a Cooper pair splitter has been demonstrated by the nonlocal conductance \cite{Hofstetter,Herrmann} and noise \cite{Wei,Das} measurement. The finite bias Cooper pair splitting \cite{Hofstetter2} and high purity of nonlocal transport by CAR \cite{Schindele} were also reported in the quantum dots based splitter. However, the main limitation of the existing proposals is that the entanglement production rate is still at low level, even when the competitive elastic cotunneling (EC) process is filtered out \cite{Recher,Feinberg,Linder,Linder2,Cayssol,Veldhorst,Chen,Yeyati,Golubev,Herrera,Chen,Beenakker,Reinthaler}. In order to manipulate and probe entangled states \cite{Bell1,Bell2,Clauser,Kawabata,Chtchelkatchev,Samuelsson,Samuelsson2,Beenakker2,Beenakker3,Chen3,Braunecker,Burkard,Chen4,Chen5}, the entanglement production rate demands well improvement. Another constraint on a conventional Cooper pair splitter is that the superconductor size $L$, the distance between two normal metal-superconductor (NS) interfaces, cannot strongly exceed the superconducting coherence length $\xi_0=\hbar v_F/\Delta$ \cite{Falci}. A solution  of such a  geometric constraint will greatly facilitate the fabrication of the splitter.

\begin{figure}
\centering
\includegraphics[width=0.5\textwidth]{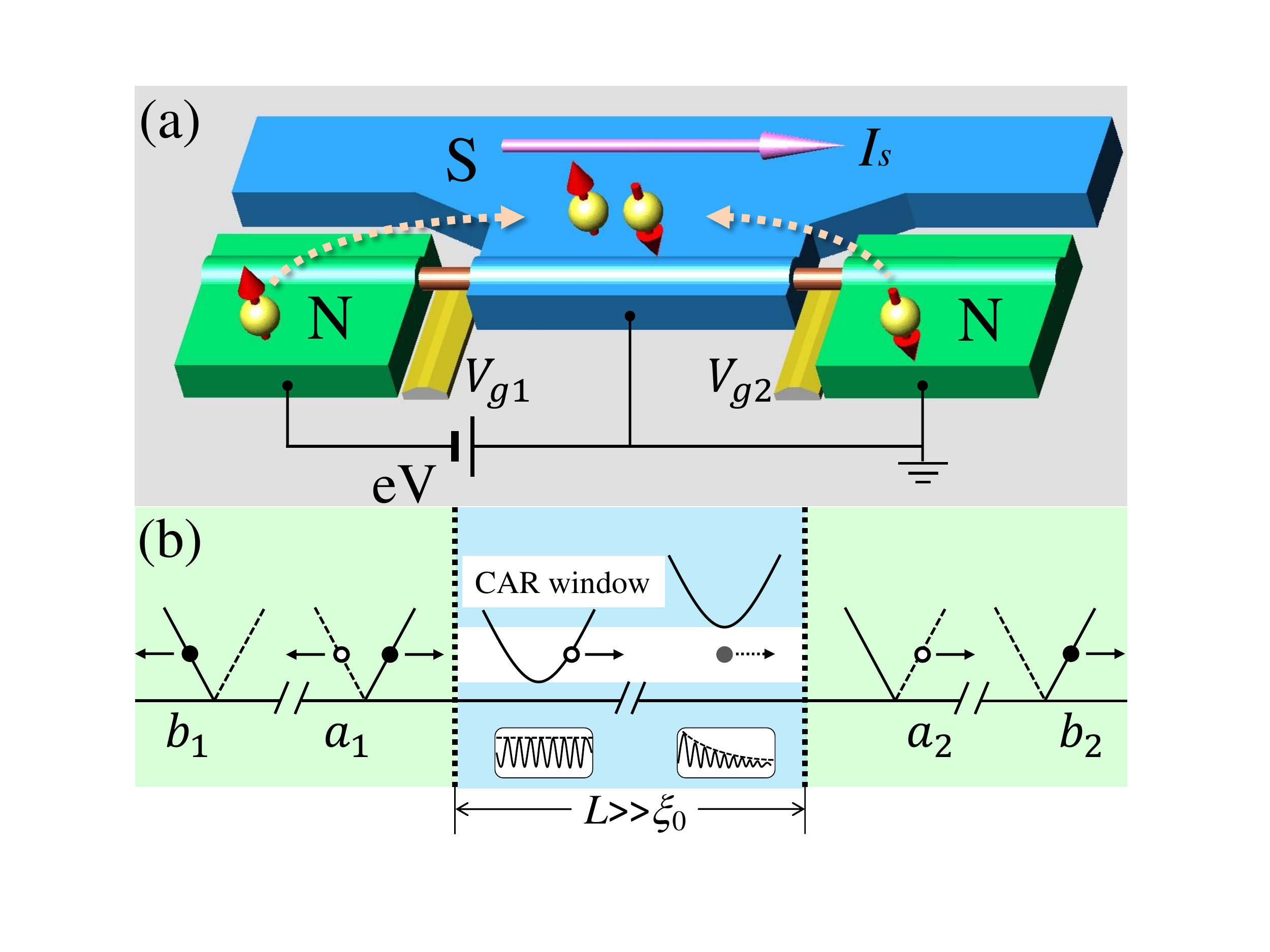}
\caption{(a) Illustration of the NSN junction fabricated on a nanowire. A supercurrent $I_s$ sketched by the long pink arrow is driven in the S. Two electrons (yellow balls with their spins labeled by the red arrows) from different N regions enter into S and form a Cooper pair during the CAR process. Two gates $V_{g1,2}$ (the golden bars) are located at the interfaces. (b) The quasiparticle picture of CAR, with the filled (open) circles representing the electron-like (hole-like) quasiparticles. The supercurrent opens a CAR window sketched by the blank region. The free wave of the hole-like quasiparticle and the evanescent wave of the electron-like quasiparticle are sketched by the inserted boxes.} \label{fig1}
\end{figure}

In this paper, we show that both the problems can be solved with the use of  a three-terminal NSN junction by driving a supercurrent in the S along the junction, as sketched in Fig. \ref{fig1}(a). The supercurrent results in opposite shifts of energy gaps for the electron-like and hole-like quasiparticles, as shown in Fig. \ref{fig1}(b). When an electron is incident to the S within the energy window (referred as the CAR window) between the two modified gaps, the hole-like quasiparticle is able to propagate freely in the S, supporting a long-range CAR, while the electron-like quasiparticle decays as usual, leading to a suppression of EC. Therefore, the probability of CAR gets significantly enhanced within the CAR window and oscillates with $L$. By investigating the entangled states via the inverse CAR process non-perturbatively, we find that there exist two types of spin singlet states, with the entangled electron pair possessing either opposite or equal energies relative to the chemical potential of the S. The total entanglement production rate depends solely on the CAR probability and a high rate close to its saturation value can be achieved.

\vspace{3ex}

\noindent
\textbf{Long-range CAR}

\noindent
To be specific, we first analyze the CAR in a nanowire NSN junction as shown in Fig. \ref{fig1}(a), where an effective pair potential is induced in the S region due to the proximity effect of the s-wave superconductor. When a supercurrent $I_s$ is driven in the S, the effective order parameter in the nanowire takes the form of $\Delta e^{2iqx}$ \cite{Romito,Chen2}. The Cooper pair momentum $2q$ can be tuned by the supercurrent through $q=i_s/\xi_0$, with  $i_s=I_s/I_c$ normalized by its critical value $I_c$. The whole system can be well described by the  Bogoliubov-de Gennes Hamiltonian as
\begin{equation}\label{BdG}
\mathcal{H}=\left(
              \begin{array}{cc}
                -\frac{\hbar^2}{2m}\partial_x^2-\mu+U(x) & i\sigma_y\Delta(x) \\
                -i\sigma_y\Delta^*(x) & \frac{\hbar^2}{2m}\partial_x^2+\mu-U(x) \\
              \end{array}
            \right),
\end{equation}
where $\sigma_y$ is the spin Pauli matrix, $\mu$ is the chemical potential, and the pair potential is given  by  $\Delta(x)=\Delta e^{2iqx}\Theta(x)\Theta(L-x)$ with $\Theta(x)$ the Heaviside step function. The Dirac-type interface potentials $U(x)=U_1\delta(x)+U_2\delta(x-L)$ are introduced to model the barriers at the NS interfaces \cite{Blonder}, which can be tuned by gates $V_{g1,2}$ as shown in Fig. \ref{fig1}(a).

Hamiltonian (\ref{BdG}) in the S region can be diagonalized by assuming the wave function as $(ue^{i(k+q)x}, ve^{i(k-q)x})^\text{T}$. The excitation energy around $\pm k_F$ can be obtained as $E=\pm i_s\Delta+\sqrt{(\hbar v_F\delta k_\pm)^2+\Delta^2}$ under the conditions of $q\ll k_F$ and $\Delta\ll\mu$, with the small wave vectors denoted by $\delta k_\pm=k\mp k_F$. The energy spectra indicate that the excitation gaps around Fermi points $\pm k_F$ shift by opposite values of $\pm i_s\Delta$, as illustrated in Fig. \ref{fig1}(b). Such energy splitting opens a CAR window, $E/\Delta\in(1-i_s,1+i_s)$, which provides an opportunity to enhance the CAR by filtering out the EC. This can be understood by solving the scattering problem of a spin-up electron incident from the left N lead into the NSN junction. The wave functions in the three regions are given by
\begin{equation}\label{wave}
\begin{split}
\Psi_\text{N}^\text{L}=&\left(
                   \begin{array}{c}
                     1 \\
                     0 \\
                   \end{array}
                 \right)e^{ik_Fx}
                 +a_1\left(
                       \begin{array}{c}
                         0 \\
                         1 \\
                       \end{array}
                     \right)
                 e^{ik_Fx}
                 +b_1\left(
                       \begin{array}{c}
                         1 \\
                         0 \\
                       \end{array}
                     \right)
                 e^{-ik_Fx},\\
\Psi_\text{N}^\text{R}=&b_2\left(
                      \begin{array}{c}
                        1 \\
                        0 \\
                      \end{array}
                    \right)
e^{ik_Fx}+a_2\left(
               \begin{array}{c}
                 0 \\
                 1 \\
               \end{array}
             \right)
e^{-ik_Fx},\\
\Psi_\text{S}=&\sum_{\tau=\pm} \lambda_\tau \left(
                                                                       \begin{array}{c}
                                                                         u_\tau \\
                                                                         v_\tau \\
                                                                       \end{array}
                                                                     \right)
e^{\tau ik_F(1+\beta_\tau)x}
+\chi_\tau\left(
                                       \begin{array}{c}
                                         v_\tau \\
                                         u_\tau \\
                                       \end{array}
                                     \right)
e^{\tau ik_F(1-\beta_\tau)x},
\end{split}
\end{equation}
where the electron and hole wave components around $\pm k_F$ are $u_\pm=\frac{1}{\sqrt{2}}[1+\sqrt{\omega_\pm^2-1}/\omega_\pm]^{1/2}$ with $\omega_\pm =E/\Delta\mp i_s$ and $v_\pm=\sqrt{1-u_\pm^2}$, respectively, and $\beta_\pm=\Delta\sqrt{\omega_\pm^2-1}/(2\mu)$. The scattering amplitudes $a_1$, $a_2$, $b_1$, and $b_2$ denote the AR, CAR, normal reflection, and EC, respectively, as shown in Fig. \ref{fig1}(b).

\begin{figure}
\centering
\includegraphics[width=0.8\textwidth]{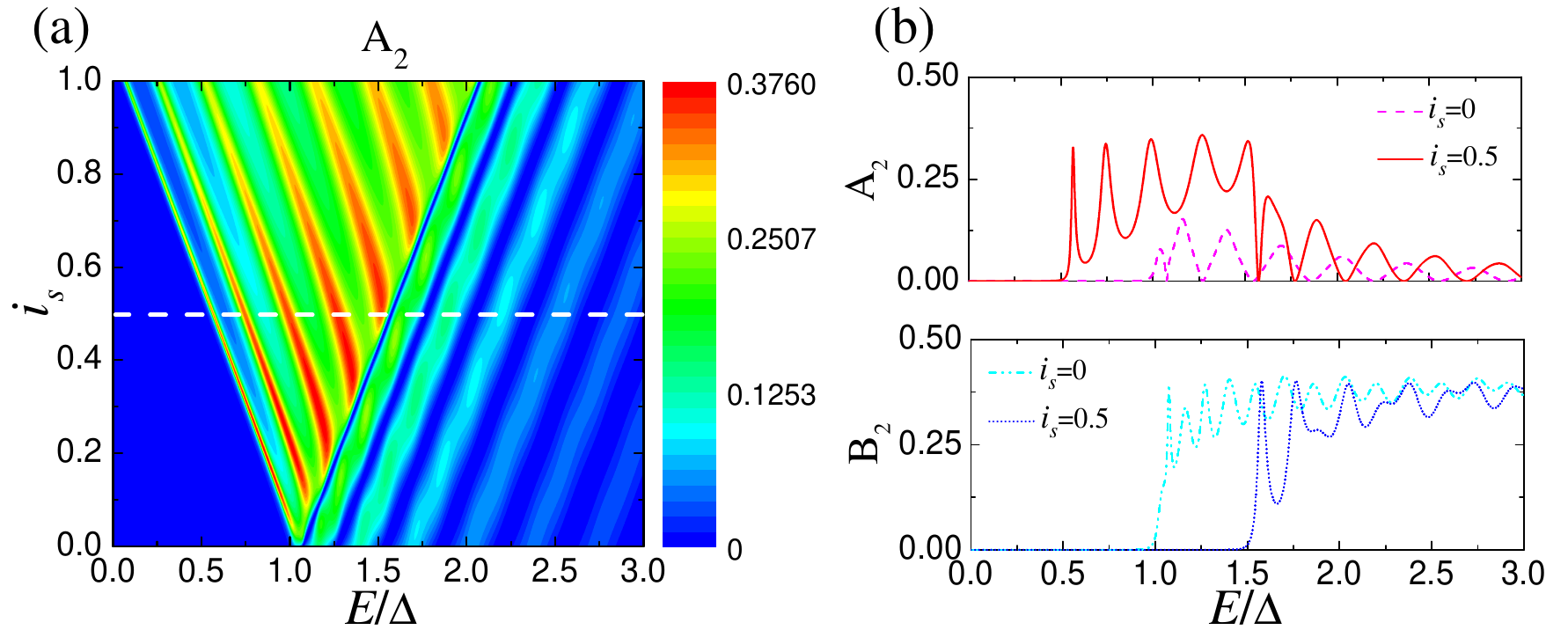}
\caption{(a) The probability of CAR as a function of supercurrent $i_s$ and energy $E$. (b) The probabilities of CAR and EC in the absence and presence of the supercurrent. The relevant parameters are set as $L=8\xi_0, Z_1=1.25, Z_2=0,$  and $k_F=100/\xi_0$.} \label{fig2}
\end{figure}

All the scattering amplitudes are solved through the boundary conditions of $\Psi_\text{N}=\Psi_\text{S}$ and $\partial_x\Psi_\text{S}-\partial_x\Psi_\text{N}=\pm2k_F Z_j\Psi_\text{S}$ at two NS interfaces, with the $``\pm"$ and the index $j(=1,2)$ corresponding to the interfaces at $x=0$ and $L$, respectively. The dimensionless barrier strength is defined as $Z_j=mU_j/(\hbar^2k_F)$. As the right NS interface is transparent  ($Z_2 =0$), the CAR amplitude takes a simple form as
\begin{equation}\label{a2}
a_2=\frac{i\sin\varphi_2f(\varphi_1)Z_1}{f(\varphi_1)f(\varphi_2)Z_1^2-g(\varphi_1)g(\varphi_2)(1+Z_1^2)},
\end{equation}
where the auxiliary functions are defined by $f(x)=\sin(il\sin x)$ and $ g(x)=\sin(il\sin x-x)$ with  $l=L/\xi_0$, and the energy dependent phases are defined by $\varphi_{1,2}=\cos^{-1}\omega_\pm$ for $\omega_\pm\leq1$ and $\varphi_{1,2}=-i\cosh^{-1}\omega_\pm$ for $\omega_\pm>1$.

As $E$ is within the CAR window, exponential factors $\beta_+$ and $\beta_-$ in Eq. (\ref{wave}) take imaginary and real values, respectively,  corresponding to an evanescent wave of the electron-like quasiparticle in the $+k_F$ branch and a free wave of the hole-like quasiparticle in the $-k_F$ branch. For $Z_2 =0$, since there is no branch-crossing scattering  at the right interface, propagations of the electron-like and hole-like quasiparticles directly contribute to the EC and CAR processes, respectively. In this case, there exists only the CAR process in the long-range limit $L\gg\xi_0$. The numerical results of the CAR probability $A_2=|a_2|^2$ as a function of $i_s$ and $E$ is plotted in Fig. \ref{fig2}(a). One can see that there is a notable region confined by the boundary approximately described by $E/\Delta=1\pm i_s$, where the CAR gets effectively enhanced. At the resonant energy levels, the CAR probability can reach a high value of 38\%. In Fig. \ref{fig2}(b), we compare the CAR probability $A_2$ and EC probability $B_2=|b_2|^2$ in the absence and presence of  the supercurrent. For a usual NSN junction of $i_s=0$, both CAR and EC processes are suppressed within the initial gap $\Delta$ due to the subgap decay of the normal and anomalous propagators in the S region, while EC dominates the nonlocal transport above the gap. Impressively, when a supercurrent $i_s=0.5$ is driven, the long-range CAR occurs within the CAR window $E/\Delta\in(0.5,1.5)$, with the  EC being quenched below the modified gap $1.5\Delta$. This is a new kind of energy filtering effect, which occurs in the S, in contrast to the previous proposal in which the energy filtering is enforced in the normal leads and a large mismatch of Fermi velocities is inevitable \cite{Cayssol,Veldhorst}.

\begin{figure}
\centering
\includegraphics[width=0.8\textwidth]{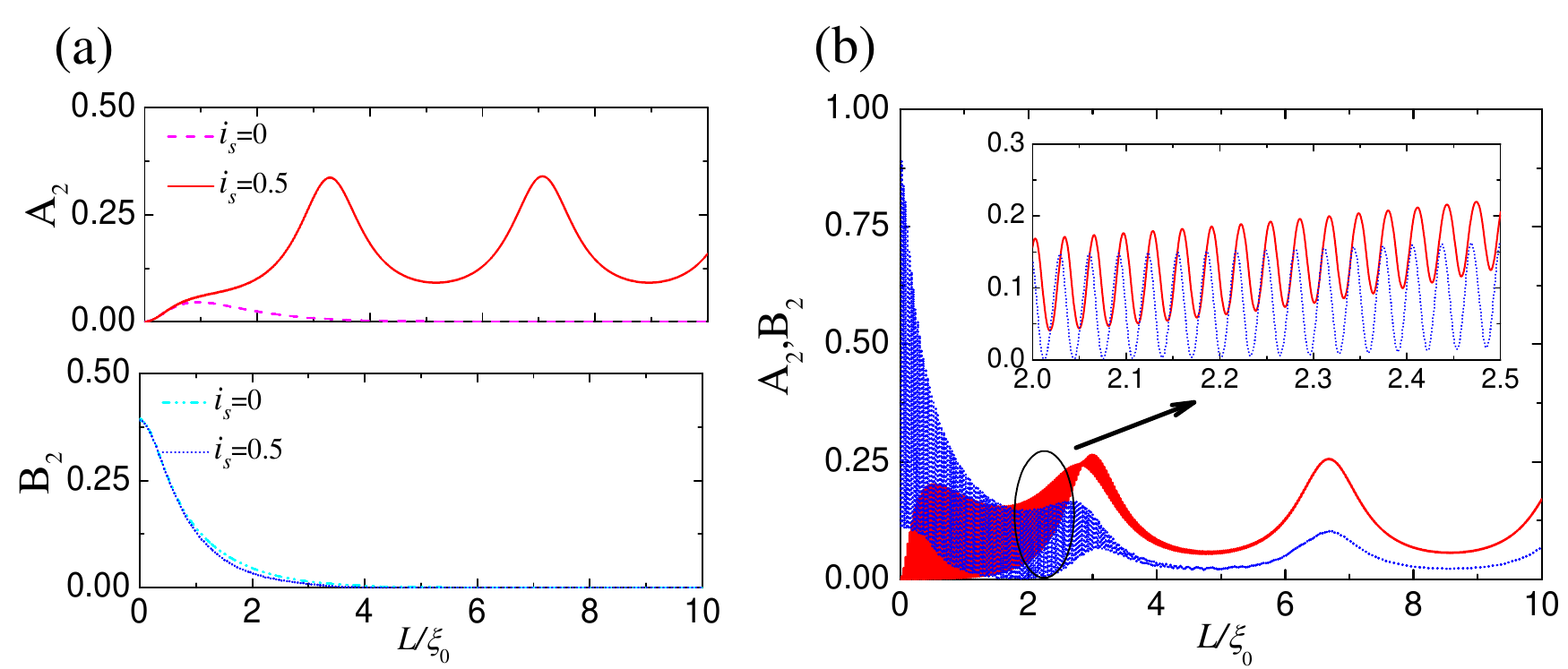}
\caption{The probabilities of CAR and EC as functions of $L$ for (a) $Z_2=0$ and (b) $Z_2=0.8$. The relevant parameters are set as $Z_1=1.25, E/\Delta=0.8, k_F=100/\xi_0$.} \label{fig3}
\end{figure}

Next, we investigate the anomalous dependence of the CAR probability on $L$ of the S region. Within the CAR window, $A_2$ and $B_2$ as functions of $L$ are presented in Fig. \ref{fig3}(a). In the absence of supercurrent, $A_2$ first increases with $L$ and then rapidly decreases, while $B_2$ monotonically decays with $L$, as known in the usual case. As a supercurrent is driven, while the result of $B_2$ changes little, the situation for $A_2$  is quite different. With increasing $L$, $A_2$ first increases  and then exhibits an oscillatory behavior due to the interference between two NS interfaces. In the large $L$ limit,  we have $g(\varphi_1)/f(\varphi_1)\simeq e^{i\varphi_1}$ within the CAR window, so that the CAR amplitude in Eq. (\ref{a2}) reduces to $a_2=i\sin\varphi_2 Z_1/[f(\varphi_2)Z_1^2-e^{i\varphi_1}g(\varphi_2)(1+Z_1^2)]$, which corresponds to an oscillation period of $L_0=\pi\xi_0/\sinh(i\varphi_2)$. Under the condition of $Z_2=0$, the scattering between the $\pm k_F$ branches at the right NS interface is absent, so that the oscillation period is on a scale of $\xi_0$. The result  for  finite $Z_2$ is shown in Fig. \ref{fig3}(b). In this case, the branch-crossing scattering  can take place at the right NS interface, so that for $L\sim\xi_0$, $A_2$ and $B_2$ exhibit  fast oscillations with the period comparable with $1/k_F$. As $L$ gets larger, the propagation of quasiparticles around $+k_F$ branch cannot survive after walking such a long distance. Therefore, the fast oscillation disappears  and the interference is contributed  again  by the free waves around $-k_F$. Due to the branch-crossing scattering at the right NS interface, the hole-like quasiparticle can now be transferred into an electron in the right lead, resulting in a finite EC probability. However, it can be shown that the ratio of the EC probability to the CAR one approximates to $B_2/A_2=Z_2^2/(1+Z_2^2)$ for an NSN junction of large $L$, so that the nonlocal transport is always dominated by the CAR for small $Z_2$. On the other hand, in the tunneling limit of  $Z_2\gg1$, the two nonlocal processes cancel each other out, consistent with the conventional result \cite{Falci}.

By applying a negative bias voltage $eV\in\Delta(1-i_s,1+i_s)$ on the left lead while keeping the S and the right lead grounded as shown in Fig. \ref{fig1}(a), the nonlocal differential conductance within the CAR window is equal to $G_2/G_0=A_2/(1+Z_2^2)$ with $G_0=e^2/h$ the unit conductance (the positive direction of current is from the right lead into the S). The positive nonlocal conductance manifests a CAR-dominant nonlocal transport. The dependence of $Z_2$ on the gate voltage $V_{g2}$ can be calibrated through the normal state transport beforehand, so that the CAR probability can be measured directly by conductance $G_2$.

\vspace{3ex}

\noindent
\textbf{Entanglement generation via inverse CAR}

\noindent
By imposing a positive bias voltage ($eV<0$) on the left lead and reversing the direction of the supercurrent, the Cooper pair can break up into two nonlocal spin entangled electrons. For the high-efficiency inverse CAR, the entangled state should be analyzed non-perturbatively.  We start with the many-body state of  incident holes occupying the energy window from the Fermi level to $|eV|$ in the left N region as $|\Psi_{\text{in}}\rangle=\prod_{0<E<|eV|}\gamma_{L\uparrow,E}^{\text{i}\dag}\gamma_{L\downarrow,E}^{\text{i}\dag}|0\rangle$,
where $\gamma_{L\sigma,E}^{\text{i}\dag}$ generates an incident hole in the left lead with energy $E$ and spin $\sigma(=\uparrow,\downarrow)$, and the vacuum state $|0\rangle$ represents the Fermi sea filled up to $E=0$ in all NSN regions. The incident and outgoing waves are related to each other  via scattering coefficients as \cite{Samuelsson,Beenakker2}
\begin{equation}\label{scatter}
\gamma_{L\sigma}^{\text{i}\dag}=b_{1\sigma}^{h}\gamma_{L\sigma}^{\text{o}\dag}+a_{1\sigma}^hc_{L\sigma}^{\text{o}\dag}
+b_{2\sigma}^h\gamma_{R\sigma}^{\text{o}\dag}+a_{2\sigma}^hc_{R\sigma}^{\text{o}\dag},
\end{equation}
where $\gamma_{\alpha\sigma}^{\text{o}\dag}$ and $c_{\alpha\sigma}^{\text{o}\dag}$ are  the operators of the outgoing hole and electron in lead $\alpha(=L,R)$, respectively, with the energy index omitted, and the scattering amplitudes correspond to those in Eq. (\ref{wave}), except that the superscript denote the hole incident case. Substituting Eq. (\ref{scatter}) into the expression of $|\Psi_\text{in}\rangle$, one arrives at the outgoing state of quasiparticles. Then the entangled state between electrons can be obtained by redefining a new vacuum state $|\tilde{0}\rangle$, which is related to the original one through $|0\rangle=\prod_{eV<E<0}c_{L\uparrow,E}^{\text{o}\dag}c_{L\downarrow,E}^{\text{o}\dag}|\tilde{0}\rangle$, and by performing a particle-hole transformation as $\gamma_{\alpha\sigma,E}^{\text{o}\dag}=c_{\alpha\bar{\sigma},-E}^{\text{o}}$ \cite{Samuelsson,Prada}. The many-body outgoing state can be obtained as
\begin{equation}\label{entangle}
\begin{split}
&|\Psi_\text{out}\rangle=\prod\limits_{0<E<|eV|}\left(\kappa|\mathcal{E}\rangle+\tilde{\kappa}|\tilde{\mathcal{E}}\rangle+|\mathcal{O}\rangle\right),\\
&|\mathcal{E}\rangle=\frac{\sqrt{2}}{2}\big(c_{R\uparrow,E}^{\text{o}\dag}c_{L\downarrow,-E}^{\text{o}\dag}-c_{R\downarrow,E}^{\text{o}\dag}c_{L\uparrow,-E}^{\text{o}\dag}\big)|\tilde{0}\rangle,\\
&|\tilde{\mathcal{E}}\rangle=\frac{\sqrt{2}}{2}\big(c_{R\uparrow,E}^{\text{o}\dag}c_{L\downarrow,E}^{\text{o}\dag}-c_{R\downarrow,E}^{\text{o}\dag}c_{L\uparrow,E}^{\text{o}\dag}\big)c_{L\downarrow,-E}^{\text{o}\dag}c_{L\uparrow,-E}^{\text{o}\dag}|\tilde{0}\rangle,
\end{split}
\end{equation}
where $|\mathcal{E}\rangle$ and $|\tilde{\mathcal{E}}\rangle$ are two kinds of nonlocal spin singlet states induced by the inverse CAR with amplitude $\kappa=-\sqrt{2}a_2^*(-E)b_1^*(-E)$ and $\tilde{\kappa}=\sqrt{2}a_1^*(-E)a_2^*(-E)$, respectively, and $|\mathcal{O}\rangle$ includes the other many-body scattering states, such as the local entangled state and product states (Supplementary information). During the derivation of Eq. (\ref{entangle}), we have taken into account the particle-hole symmetry of the scattering matrix, and expressed the state with the scattering amplitudes in the electron incident case.

\begin{figure}
\centering
\includegraphics[width=0.6\textwidth]{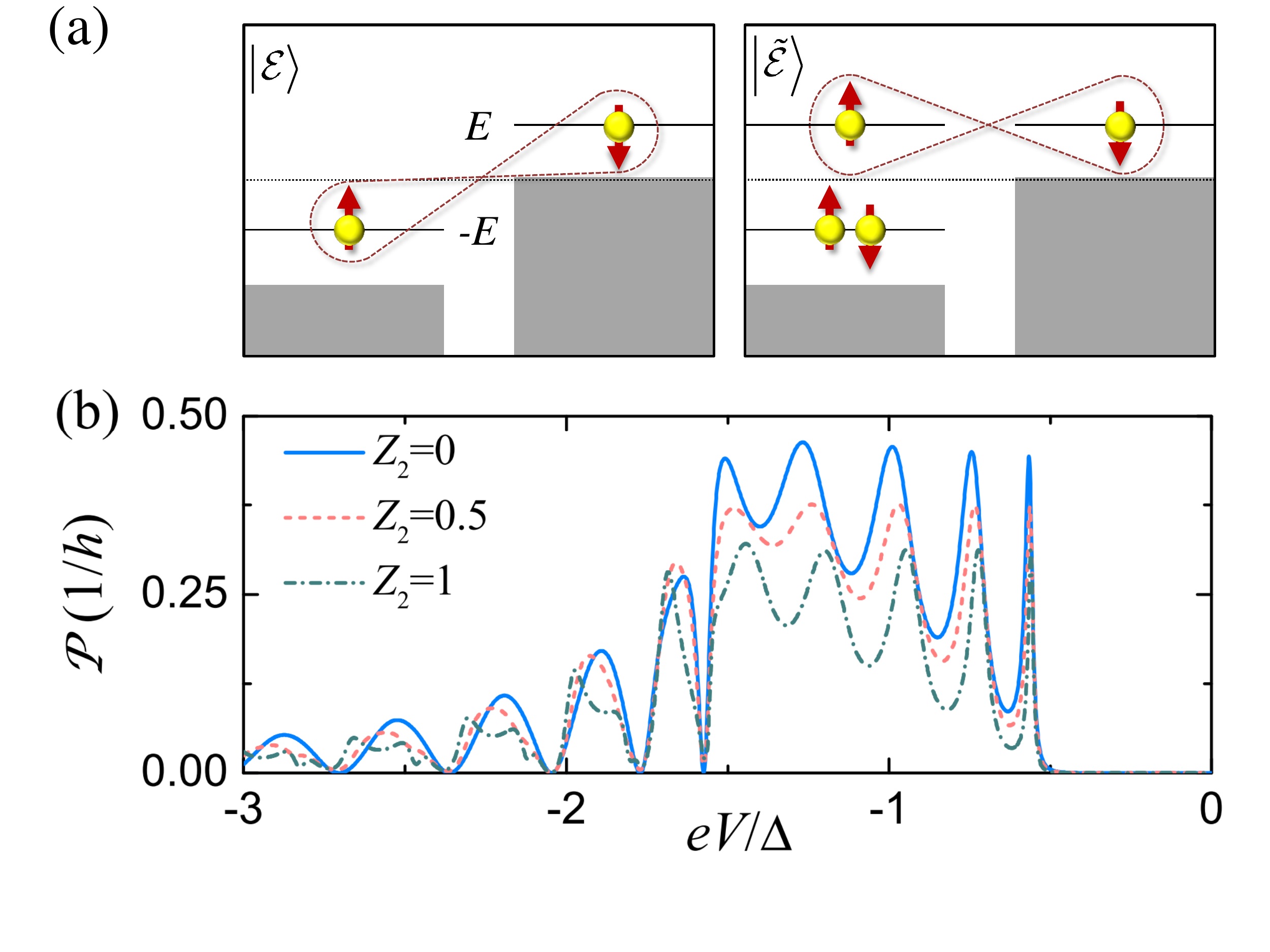}
\caption{(a) Schematic of entangled states between opposite and equal energy levels. (b) Entanglement production rate $\mathcal{P}$ as a function of bias voltage for different $Z_2$. The relevant parameters are set as $L=8\xi_0, Z_1=1.25, i_s=-0.5, k_F=100/\xi_0$.} \label{fig4}
\end{figure}

$|\mathcal{E}\rangle$ in Eq.~(5) is the usual nonlocal entangled state generated by the CAR, for which the two entangled electrons have opposite energies relative to $\mu$ \cite{Samuelsson,Prada}, as shown in the left panel of Fig. \ref{fig4}(a). Since the CAR is an elastic process, the total energy of the two electrons is conserved during the pair breaking. We note that the amplitude $\kappa$ reduces to $\sqrt{2}a^*_2$ in the tunneling limit of $b^*_1\approx1$, consistent with the result in previous literatures \cite{Samuelsson,Prada}.

$|\tilde{\mathcal{E}}\rangle$ is the anomalous entangled state in which the two entangled electrons have the same energy $E$,  as shown in the right panel of Fig. \ref{fig4}(a), which is a natural result obtained in the  present non-perterbative approach. From a physical point of view,  the coincidence of  the AR and CAR processes first results in a double electron occupation at the $-E$ energy level, freezing out the spin freedom therein; then
a nonlocal entangled state is induced for the electron pair with equal energy $E$ and opposite spins.
Such a nonlocal  entangled state has never been discussed before, since in the previous consideration,  both $a_1$ and $a_2$ are very small in the tunneling regime, and  $\tilde{\kappa}$ becomes a negligible higher-order contribution \cite{Samuelsson,Prada}.
In the present proposal, neither $a_1$ nor $a_2$ is small, so that the state $|\tilde{\mathcal{E}}\rangle$ becomes observable.

Given that tunnel attempts in a unit energy interval occur with a frequency $1/h$ ($h$ is the Plank constant) \cite{Beenakker3,Levitov}, the differential production rate of nonlocal entanglement can be calculated by $\mathcal{P}=(|\kappa|^2+|\tilde{\kappa}|^2)/h$. By further utilizing the current conservation of the quasiparticles, we obtain
\begin{equation}
\mathcal{P}(eV)=2 A_2\left(1-\frac{1+2Z_2^2}{1+Z_2^2}A_2\right)/h,
\end{equation}
in the CAR window for the Cooper pair splitter, which is determined  solely by the CAR probability.  In the case of $Z_2=0$, it is simplified to $\mathcal{P}=2(1-A_2)A_2/h$, indicating that a saturation value $\mathcal{P}=0.5/h$ can be achieved at $A_2=0.5$. It turns out that the entanglement production rate $\mathcal{P}$ is \emph{not} a monotone increasing function of the CAR probability.  Only in the tunnel limit,
$\mathcal{P}$ increases monotonously with $A_2$ because of $A_2\ll 1$.
The numerical result of $\mathcal{P}$ as a function of the bias voltage $eV$ is shown in Fig. \ref{fig4}(b). When the right NS interface is transparent, $\mathcal{P}$ can approach its saturation value at the resonant energy levels within the CAR window, indicating that an ideal entangler can be implemented in our proposal.  Although finite $Z_2$ may suppress the entanglement production rate, large values of $\mathcal{P}$ can be obtained even for a stronger barrier $Z_2=1$, as shown in Fig. \ref{fig4}(b).

The nonlocal spin entanglement of states $|\mathcal{E}\rangle$ and $\tilde{|\mathcal{E}\rangle}$ can be demonstrated through either the violation of the Bell-inequality \cite{Kawabata,Chtchelkatchev,Samuelsson,Samuelsson2,Beenakker2,Beenakker3,Chen3,Braunecker} or a well-designed spintronic quantum eraser \cite{Chen4,Chen5}. Both schemes can be achieved by the spin resolved current correlation measurement. In the optimal case of  $Z_2=0$, the nonlocal spin correlation is purely contributed by $|\mathcal{E}\rangle$ and $|\tilde{\mathcal{E}}\rangle$. In the case of  $Z_2\neq0$, the signal of entanglement will get weakened since the intervention of EC contributes an opposite current; nevertheless, provided that  the CAR dominates the nonlocal transport in the CAR window, the signal of entanglement is always extractable. Moreover, the equal-energy entangled state $\tilde{|\mathcal{E}\rangle}$ can also be probed by the bunching behavior in a beam splitter setup \cite{Burkard}, in consideration of  the orbital wave function of the singlet state being symmetric.

\vspace{3ex}

\noindent
\textbf{Discussions}

\noindent
Finally, we discuss the experimental realization of our proposal. The long-range Cooper pair splitter  in Fig. \ref{fig1}(a) possesses the same configuration as those in the experiments reported in Refs. \cite{Hofstetter,Schindele,Das,Herrmann}, so that it can be implemented by the existing technology. Since no quantum dot embedded in the nanowire needs to be fabricated, more kinds of nanowires can be adopted to build the NSN junction. A key point is that there needs to be  a suppercurrent flowing along the junction, which can be driven by imposing a constant current on the S or by utilizing a magnetic flux in a superconducting loop. Possible renormalization of the Fermi velocity due to the superconducting proximity effect can be compensated by a back gate. In order to realize an entangler with high efficiency and extract pure signal of entanglement, the interface barriers $Z_{1,2}$ must be carefully tuned by gate voltages  $V_{g1,2}$. The optimal choice  corresponding to the bias voltage configuration in Fig. \ref{fig1}(a) is  found to be $Z_1\in(1,2)$ and $Z_2=0$, which can be realized in the experiment \cite{Das}.

\noindent
\textbf{Acknowledgments}

\noindent
We would like to thank A. Baumgartner, M. Heiblum, Jing-Min Hou, Ming Gong, Zheng-Yuan Xue and Tao Zhou for helpful discussions. This work is supported by the State Key Program for
Basic Researches of China under Grant No. 2011CB922103, and by the National Natural Science
Foundation of China under Grants No. 11174125, No. 91021003 and No. 11374159.

\noindent
\textbf{Author contributions}

\noindent
W.C. conceived the project and performed the calculations. W.C. and D.Y.X. wrote the manuscript. All authors discussed the results and reviewed the manuscript.

\noindent
\textbf{Additional information}

\noindent
Supplementary information accompanies this paper

\noindent
Competing financial interests: The authors declare no competing financial interests.

\noindent
\textbf{Figure Legends}

Fig.1: (a) Illustration of the NSN junction fabricated on a nanowire. A supercurrent $I_s$ sketched by the long pink arrow is driven in the S. Two electrons (yellow balls with their spins labeled by the red arrows) from different N regions enter into S and form a Cooper pair during the CAR process. Two gates $V_{g1,2}$ (the golden bars) are located at the interfaces. (b) The quasiparticle picture of CAR, with the filled (open) circles representing the electron-like (hole-like) quasiparticles. The supercurrent opens a CAR window sketched by the blank region. The free wave of the hole-like quasiparticle and the evanescent wave of the electron-like quasiparticle are sketched by the inserted boxes.

Fig.2: (a) The probability of CAR as a function of supercurrent $i_s$ and energy $E$. (b) The probabilities of CAR and EC in the absence and presence of the supercurrent. The relevant parameters are set as $L=8\xi_0, Z_1=1.25, Z_2=0,$  and $k_F=100/\xi_0$.

Fig.3: The probabilities of CAR and EC as functions of $L$ for (a) $Z_2=0$ and (b) $Z_2=0.8$. The relevant parameters are set as $Z_1=1.25, E/\Delta=0.8, k_F=100/\xi_0$.

Fig.4: (a) Schematic of entangled states between opposite and equal energy levels. (b) Entanglement production rate $\mathcal{P}$ as a function of bias voltage for different $Z_2$. The relevant parameters are set as $L=8\xi_0, Z_1=1.25, i_s=-0.5, k_F=100/\xi_0$.


\begin{thebibliography}{99}
\bibitem{Nielsen} Nielsen, M. A. \& Chuang, I. L. \emph{Quantum Computation and Quantum Information} (Cambridge University Press, Cambridge, England, 2000).
\bibitem{Recher} Recher, P., Sukhorukov, E. V. \& Loss, D. Andreev tunneling, Coulomb blockade, and resonant transport of nonlocal spin-entangled electrons. \emph{Phys. Rev. B} \textbf{63}, 165314 (2001).
\bibitem{Lesovik} Lesovik, G. B., Martin, T. \& Blatter, G. Electronic entanglement in the vicinity of a superconductor. \emph{Eur. Phys. J. B} \textbf{24}, 287 (2001).
\bibitem{Byers} Byers, J. M. \& Flatt\'{e}, M. E. Probing Spatial Correlations with Nanoscale Two-Contact Tunneling. \emph{Phys. Rev. Lett.} \textbf{74}, 306 (1995).
\bibitem{Feinberg} Deutscher, G. \& Feinberg, D. Coupling superconducting-ferromagnetic point contacts by Andreev reflections. \emph{Appl. Phys. Lett.} \textbf{76}, 487 (2000).
\bibitem{Hofstetter} Hofstetter, L., Csonka, S., Nyg{\aa}rd, J. \& Sch\"{o}nenberger, C. Cooper pair splitter realized in a two-quantum-dot Y-junction. \emph{Nature}  (London) \textbf{461}, 960 (2009).
\bibitem{Herrmann} Herrmann, L. G., Portier, F., Roche, P., Yeyati, A. L., Kontos, T. \& Strunk, C. Carbon Nanotubes as Cooper-Pair Beam Splitters. \emph{Phys. Rev. Lett. } \textbf{104}, 026801 (2010).
\bibitem{Wei} Wei, J. \& Chandrasekhar, V. Positive noise cross-correlation in hybrid superconducting and normal-metal three-terminal devices. \emph{Nat. Phys.} \textbf{6}, 494 (2010).
\bibitem{Das} Das, A., Ronen, Y., Heiblum, M., Mahalu, D., Kretinin, A. V. \& Shtrikman H. High-efficiency Cooper pair splitting demonstrated by two-particle conductance resonance and positive noise cross-correlation. \emph{Nat. Commun.} \textbf{3}, 1165 (2012).
\bibitem{Hofstetter2} Hofstetter, L., Csonka, S., Baumgartner, A., F\"{u}l\"{o}p, G., d'Hollosy, S., Nyg{\aa}rd, J. \& Sch\"{o}nenberger, C. Finite-Bias Cooper Pair Splitting. \emph{Phys. Rev. Lett.} \textbf{107}, 136801 (2011).
\bibitem{Schindele} Schindele, J., Baumgartner, A. \& Sch\"{o}nenberger, C. Near-Unity Cooper Pair Splitting Efficiency. \emph{Phys. Rev. Lett.} \textbf{109}, 157002 (2012).
\bibitem{Linder} Linder, J., Zareyan, M. \& Sudb{\o}, A. Spin-switch effect from crossed Andreev reflection in superconducting graphene spin valves. \emph{Phys. Rev. B} \textbf{80}, 014513 (2009).
\bibitem{Linder2} Linder, J. \& Yokoyama, T. Superconducting proximity effect in silicene: Spin-valley-polarized Andreev reflection, nonlocal transport, and supercurrent. \emph{Phys. Rev. B} \textbf{89}, 020504(R) (2014).
\bibitem{Cayssol} Cayssol, J. Crossed Andreev Reflection in a Graphene Bipolar Transistor. \emph{Phys. Rev. Lett.} \textbf{100}, 147001 (2008).
\bibitem{Veldhorst} Veldhorst, M. \& Brinkman, A. Nonlocal Cooper Pair Splitting in a pSn Junction. \emph{Phys. Rev. Lett.} \textbf{105}, 107002 (2010).
\bibitem{Chen} Chen, W., Shen, R., Sheng, L., Wang, B. G. \& Xing, D. Y. Resonant nonlocal Andreev reflection in a narrow quantum spin Hall system. \emph{Phys. Rev. B} \textbf{84}, 115420 (2011).
\bibitem{Yeyati} Yeyati, A. Levy, Bergeret, F. S., Mart\'{\i}n-Rodero A. \& Klapwijk, T. M. Entangled Andreev pairs and collective excitations in nanoscale superconductors. \emph{Nature Phys.} \textbf{3}, 455 (2007).
\bibitem{Golubev} Golubev, D. S. \& Zaikin, A. D. Non-local Andreev reflection under ac bias. \emph{Europhys. Lett.} \textbf{86}, 37009 (2009).
\bibitem{Herrera} Herrera, W. J., Levy Yeyati, A. \& Martin-Rodero, A. Long-range crossed Andreev reflections in high-temperature superconductors. \emph{Phys. Rev. B} \textbf{79}, 014520 (2009).
\bibitem{Beenakker} Nilsson, J., Akhmerov, A. R. \& Beenakker, C. W. J. Splitting of a Cooper Pair by a Pair of Majorana Bound States. \emph{Phys. Rev. Lett.} \textbf{101}, 120403 (2008).
\bibitem{Reinthaler} Reinthaler, R. W. Recher, P. \& Hankiewicz, E. M. Proposal for an All-Electrical Detection of Crossed Andreev Reflection in Topological Insulators. \emph{Phys. Rev. Lett.} \textbf{110}, 226802 (2013).
\bibitem{Bell1} Bell, J. S. On the Einstein Podolsky Rosen Paradox. \emph{Physics} (Long Island City, N.Y.) \textbf{1}, 195 (1964).
\bibitem{Bell2} Bell, J. S. On the Problem of Hidden Variables in Quantum Mechanics. \emph{Rev. Mod. Phys.} \textbf{38}, 447 (1966).
\bibitem{Clauser} Clauser, J. F., Horne, M. A., Shimony, A. \& Holt, R. A. Proposed Experiment to Test Local Hidden-Variable Theories. \emph{Phys. Rev. Lett.} \textbf{23}, 880 (1969).
\bibitem{Kawabata} Kawabata, S. Test of Bell's Inequality using the Spin Filter Effect in Ferromagnetic Semiconductor Microstructures. \emph{J. Phys. Soc. Jpn.} \textbf{70}, 1210 (2001).
\bibitem{Chtchelkatchev} Chtchelkatchev, N. M., Blatter, G., Lesovik, G. B. \& Martin, T. Bell inequalities and entanglement in solid-state devices. \emph{Phys. Rev. B} \textbf{66}, 161320(R) (2002).
\bibitem{Samuelsson} Samuelsson, P., Sukhorukov, E. V. \& B\"{u}ttiker, M. Orbital Entanglement and Violation of Bell Inequalities in Mesoscopic Conductors. \emph{Phys. Rev. Lett.} \textbf{91}, 157002 (2003).
\bibitem{Samuelsson2} Samuelsson, P., Sukhorukov, E. V. \& B\"{u}ttiker, M. Two-Particle Aharonov-Bohm Effect and Entanglement
in the Electronic Hanbury Brown¨CTwiss Setup. \emph{Phys. Rev. Lett.} \textbf{92}, 026805 (2004).
\bibitem{Beenakker2} Beenakker, C. W. J., Emary, C., Kindermann, M. \& van Velsen, J. L. Proposal for Production and Detection of Entangled Electron-Hole Pairs in a Degenerate Electron Gas. \emph{Phys. Rev. Lett.} \textbf{91}, 147901 (2003).
\bibitem{Beenakker3} Beenakker, C. W. J. Electron-hole entanglement in the Fermi sea. arXiv:cond-mat/0508488v3.
\bibitem{Chen3} Chen, W., Shen, R., Sheng, L., Wang, B. G.  \& Xing, D. Y. Electron Entanglement Detected by Quantum Spin Hall Systems. \emph{Phys. Rev. Lett.} \textbf{109}, 036802 (2012).
\bibitem{Braunecker} Braunecker, B., Burset, P. \& Levy Yeyati, A. Entanglement Detection from Conductance Measurements in Carbon Nanotube Cooper Pair Splitters. \emph{Phys. Rev. Lett.} \textbf{111}, 136806 (2013).
\bibitem{Chen4} Chen, W., Shen, R., Wang, Z. D., Sheng, L., Wang, B. G. \& Xing, D. Y. Quantitatively probing two-electron entanglement with a spintronic quantum eraser. \emph{Phys. Rev. B} \textbf{87}, 155308 (2013).
\bibitem{Chen5} Chen, W., Wang, Z. D., Shen R., \& Xing, D. Y. Probing spin entanglement by gate-voltage-controlled interference
of current correlation in quantum spin Hall insulators. \emph{Phys. Lett. A} \textbf{378}, 1893 (2014).
\bibitem{Burkard} Burkard, G., Loss, D. \& Sukhorukov, E. V. Noise of entangled electrons: Bunching and antibunching. \emph{Phys. Rev. B} \textbf{61}, R16303 (2000).
\bibitem{Falci} Falci, G., Feinberg, D. \& Hekking, F. W. J. Correlated tunneling into a superconductor in a multiprobe hybrid structure. \emph{Europhys. Lett.} \textbf{54}, 255 (2001).
\bibitem{Romito} Romito, A., Alicea, J., Refael, G. \& von Oppen, F. Manipulating Majorana fermions using supercurrents. \emph{Phys. Rev. B} \textbf{85}, 020502(R) (2012).
\bibitem{Chen2} Chen, W., Gong, M., Shen, R. \& Xing, D. Y. Detecting Fulde¨CFerrell superconductors by an Andreev interferometer. \emph{New J. Phys.} \textbf{16} 083024 (2014).
\bibitem{Blonder} Blonder, G. E., Tinkham, M. \& Klapwijk, T. M. Transition from metallic to tunneling regimes in superconducting microconstrictions: Excess current, charge imbalance, and supercurrent conversion. \emph{Phys. Rev. B} \textbf{25}, 4515 (1982).
\bibitem{Prada} Prada E. \& Sols, F. Entangled electron current through finite size normal-superconductor tunneling structures. \emph{Eur. Phys. J. B} \textbf{40}, 379 (2004).
\bibitem{Levitov} Levitov, L. S. \& Lesovik, G. B. Charge distribution in quantum shot noise. \emph{JETP Lett.} \textbf{58}, 230 (1993).
\end{thebibliography}
\end{document}